\documentclass[osajnl2,preprint,showpacs]{revtex4}

\newcommand {\Ket}[1]         {\ensuremath{| #1 \rangle}}

\newcommand {\KetBra}[2]      {\ensuremath{| #1 \rangle\langle #2 |}}
\usepackage{graphicx}

\begin{document}
\title{Enhanced four-wave mixing via crossover resonance \\ in cesium vapor}
\author{T. Passerat de Silans}
 \affiliation{Laborat\'orio de Superf\'icies, Departamento de F\'isica,
  Universidade Federal da Para\'iba, Cx. Postal 5086, 58051-900 Jo\~ao Pessoa, PB, Brazil}
\author{C. S. L. Goncalves}%
\affiliation{Departamento de
F\'isica, Universidade Federal de Pernambuco, Cidade
Universit\'aria, 50670-901 Recife, PE, Brazil}
\author{D. Felinto}
\affiliation{Departamento de F\'isica, Universidade Federal de Pernambuco, Cidade Universit\'aria, 50670-901 Recife, PE, Brazil}%
\author{J.W.R. Tabosa}
\affiliation{Departamento de F\'isica, Universidade Federal de Pernambuco, Cidade Universit\'aria, 50670-901 Recife, PE, Brazil}%

\begin{abstract}
We report on the observation of enhanced four-wave mixing via
crossover resonance in a Doppler broadened cesium vapor. Using a
single laser frequency, a resonant parametric process in a
double-$\Lambda$ level configuration is directly excited for a
specific velocity class. We investigate this process in different
saturation regimes and demonstrate the possibility of generating
intensity correlation and anti-correlation between the
probe and conjugate beams. A simple theoretical
model is developed that accounts qualitatively well to the
observed results.
\end{abstract}

\ocis{190.4380, 270.1670, 300.6290}

\maketitle

\section{Introduction}
Four-wave mixing (FWM) is a nonlinear optical process which has
been largely investigated along the past several decades, both in
homogeneously and inhomogeneously broadened atomic samples
\cite{LeBoiteux86, Lukin00, Tabosa97}. Since the very early
studies of FWM, the possibility to enhance the efficiency of this
nonlinear process through the use of resonant atomic or molecular
media has led to the investigation of FWM processes using different
atomic level configurations as, for instance, two-level
\cite{Lind78, Oria89}, three-level $\Lambda$ \cite{Pinard86,
Cardoso02}, and four-level double-$\Lambda$ schemes \cite{Lukin00}.
For instance, a highly efficient, low intensity, FWM signal has been
first observed in a double-$\Lambda$ level configuration by Hemmer
et al. \cite{Hemmer95}. More recently this particular configuration has
attracted much attention due to the generation of narrow-band
photon pairs in atomic ensembles via spontaneous FWM
\cite{Harris06}, as well as the efficient generation of pairs of
intense light beams showing a high degree of intensity squeezing
\cite{Lett07}. In those previous experiments usually two different
laser frequencies are employed to explore the parametric resonance
in the three- or four-level schemes. One common characteristic of
the FWM process using $\Lambda$ or double-$\Lambda$ atomic levels
schemes is the possibility of completely eliminating the resonant
absorption by the medium due to the phenomenon of Electromagnetically Induced
Transparency (EIT)\cite{Harris97, Fleischhauer05}, allowing also the control of the medium's refractive index
\cite{Shahriar98, Lukin98}.

In this work we investigate the FWM process in a sample of thermal
cesium atoms using a single laser frequency where different atomic
level configurations ($\Lambda$ and double-$\Lambda$) can
be directly excited depending where the laser frequency is tuned
inside the cesium Doppler profile. In particular, we demonstrate
that a resonant FWM process in a double-$\Lambda$ level scheme can
be observed via the usual crossover resonance. We also show
that the FWM spectrum is strongly dependent on the intensity of
the pump beams. Moreover, we demonstrate  that the generated
FWM beam can be correlated or anti-correlated to the incident
probe beam depending mainly on which configuration of atomic levels is
excited. It is worth mentioning that intensity correlation of two
laser beams in a coherently prepared atomic medium, under EIT
condition, have been observed before \cite{Garrido-Alzar03,Martinelli04,
Scully05}. Although these observed intensity correlations and
anti-correlations are essentially classical in nature, they add
new perspectives to the investigation of quantum correlated beams
in such atomic system, owing to the previous observation of
quantum correlated twin photon employing FWM in double-$\Lambda$
level schemes \cite{Harris06}.

The paper is organized as follows: In section II we describe the
experimental setup and the observed FWM spectra; Section III is
devoted to the development of a simple theoretical model based on the
density matrix formalism to calculate the FWM spectra in different
saturating regimes. We also present in this section a comparison
between the calculated and measured FWM spectra. In Section IV we
present the experimental results for intensity correlations and
anti-correlations observed for different laser frequencies,  and
present a qualitative discussion of the observed results. Finally,
in section V we present our conclusions.

\section{Four wave mixing in cesium vapor: Experiment}

\subsection{Experimental configuration}

\begin{figure}[h]
    \includegraphics[width=8.5cm]{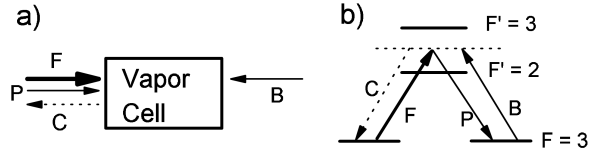}
    \caption{a) Geometry of the DBFWM experiment, beams F and P
 are copropagating and beam B is counterpropagating with respect to both.
 Beam C is generated counterpropagating with respect to beam P. b) The effective four
 level atomic system used in the experiment, with the ground levels corresponding to different
  Zeeman sublevels of the $F=3$ hyperfine level of Cs$\left(6S_{1/2}\right)$
  while the excited states are Zeeman sublevels of the hyperfine $F'=2$ and $F'=3$ levels of Cs$\left(6P_{3/2}\right)$.
  F and C have $\sigma_+$ circular polarization, while beams B and P have $\sigma_{-}$ polarization.}
    \label{fig:Config}
\end{figure}

The experiment consists in using the so called degenerated
backward four wave mixing configuration (DBFWM) in which three
beams are sent into a vapor and a fourth beam is generated by the
induced polarization. The geometry of the beams is depicted in
Fig. \ref{fig:Config}(a). Two beams, the forward pump (F) and
probe (P) beams with opposite circular polarizations, are sent
copropagating into the cell. Usually the forward pump beam is much
stronger than the probe beam. A third beam, called backward pump
(B) beam, is sent counterpropagating with respect to the formers
and has a circular polarization opposite to that of the F beam.
(we chose $\sigma_+$ for F beam and $\sigma_-$ for the P and B
beams). The beams couple to an effective four level atomic system
as depicted in Fig. \ref{fig:Config}(b), with the two degenerated
ground states corresponding to different Zeeman sublevels of the
F=3 hyperfine level of Cs$\left(6S_{1/2}\right)$ state and  the
excited states corresponding to zeeman sublevels of the hyperfine
levels F'=2 and F'=3 of Cs$\left(6P_{3/2}\right)$ state. All three
beams couple to both excited states. The conjugated beam (C) is
created by the diffraction of beam B into the Zeeman coherence
grating created by the F and P beams. Due to phase matching and
angular momentum conservation, the beam C propagates in opposite
direction to beam P and with opposite ($\sigma_+$) polarization .

\begin{figure}[h]
    \centering
        \includegraphics[width=8.5cm]{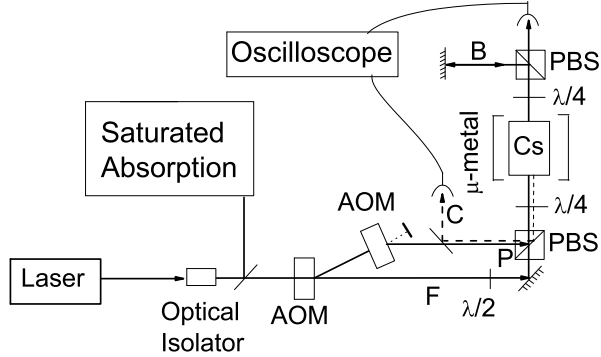}
    \caption{Experimental Setup. PBS means polarizing beam splitter and AOM stands for acousto-optical modulator.}
    \label{fig:Setup}
\end{figure}

The experimental setup is depicted in Fig. \ref{fig:Setup}. We use
an  external cavity diode laser emitting around 852 nm, with
linewidth below  1 MHz and output power of about 40 mW.  The main
beam is sent through an acousto-optical modulator, with its
non-deflected beam used as the F beam. The beam deflected in order
1 pass through another acousto-optical modulator (with the same RF
frequency as the first one) and the new deflected beam in order -1
is used as beam P. Such setup allows us to scan the frequency of
P ($\omega_P$) around the frequency of F ($\omega_F$).
The polarization of F is rotated by $90^o$ before it is combined
with P using a polarizing beam splitter (PBS). The combined
beams pass through a $\lambda/4$ plate producing the desired
circular polarizations and are then sent into the cell. The beams
diameter at the cell is around 3 mm. The
transmitted beams are converted back to linear polarization by a
second quarter wave plate and separated by another PBS. After the cube, F is retro-reflected to
produce  the beam B (with $\omega_B=\omega_F$). A 50/50 beam
splitter is used to reflect the beam C that propagates in opposite
direction to P. The laser frequency is monitored using a
saturated absorption setup.

The cell used in the four wave mixing experiment is a vacuum
sealed 12 cm long glass cell with diameter of 2.5 cm and filled
with Cs at room temperature. We have surrounded the cell with a
layer of $\mu$-metal in order to
reduce spurious effects from residual external magnetic fields.

\subsection{Experimental results}

\begin{figure}[h]
        \includegraphics[width=8.5cm]{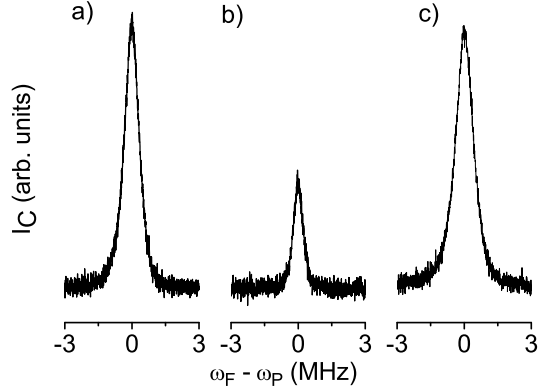}
        \caption{Intensity of the conjugated beam ($I_C$) when
        $\omega_P$  is scanned around a fixed value of $\omega_F$ for a) $\omega_F=\omega_{32}$, b) $\omega_F=\omega_{32}+2\pi\times38$ MHz, and c) $\omega_F=\omega_{CO}$. The powers used for the beams are $P_F=100\mu W$, $P_B=0.7\,P_F$, and $P_P=150\mu W$. The same vertical scale was used in all graphs.}
    \label{fig:ScanP}
\end{figure}

In Fig.~\ref{fig:ScanP} we present typical curves for the detected conjugated-beam intensity ($I_C$) as a function of $\omega_P$ for fixed values of $\omega_F$ and low pump power ($P_F=0.1\,$mW). For all our experiments, the pump power will always be given in terms of $P_F$, with $P_B = 0.7\,P_F$. Figure~\ref{fig:ScanP}(a) was obtained for $\omega_F$ resonant with the $F=3\rightarrow F'=2$ transition ($\omega_{32}$). Figure~\ref{fig:ScanP}(c) presents a similar signal obtained for $\omega_F$ at the crossover resonance between F'=2 and F'=3 ($\omega_{CO}$). For Fig.~\ref{fig:ScanP}(b), $\omega_F$ was chosen between these two values, more precisely at $\omega_F=\omega_{32}+2\pi\times38$ MHz. Note the subnatural linewidth, around 0.8~MHz, of the FWM signal in the three cases. The height of the peak, however, varies with the position of $\omega_F$ throughout the Doppler profile, with two maxima at the positions shown in Figs.~\ref{fig:ScanP}(a) and (c).

The two peaks in Figs.~\ref{fig:ScanP}(a) and (c) are associated with
different atomic level schemes: the peak at transition
$\omega_F=\omega_{32}$ corresponds mainly to all beams resonant with this
transition for atoms with zero velocity along the laser
beams, and thus to a three-level $\Lambda$ scheme; the peak at
$\omega_F=\omega_{CO}$ corresponds mainly to the case where P and F
are resonant with the transition $F=3\rightarrow F'=2$
($F=3\rightarrow F'=3$) and B is resonant with the transition
$F=3\rightarrow F'=3$ ($F=3\rightarrow F'=2$) for atoms with a
velocity component along the beam's axis of $v_z=\frac{\Delta_F}{2k}$
($v_z=-\frac{\Delta_F}{2k}$), with $\Delta_F$ the frequency
difference between the two excited levels and k the F-beam
wavenumber. Thus, for such velocity class, the system corresponds
to a double-$\Lambda$ level scheme. In our experiment, we have then a system that
allows us to change the effective experimental atomic level configuration
from $\Lambda$ to a double-$\Lambda$ by simply tuning
finely the laser frequency.

\begin{figure}[h]
    \centering
        \includegraphics[width=8.5cm]{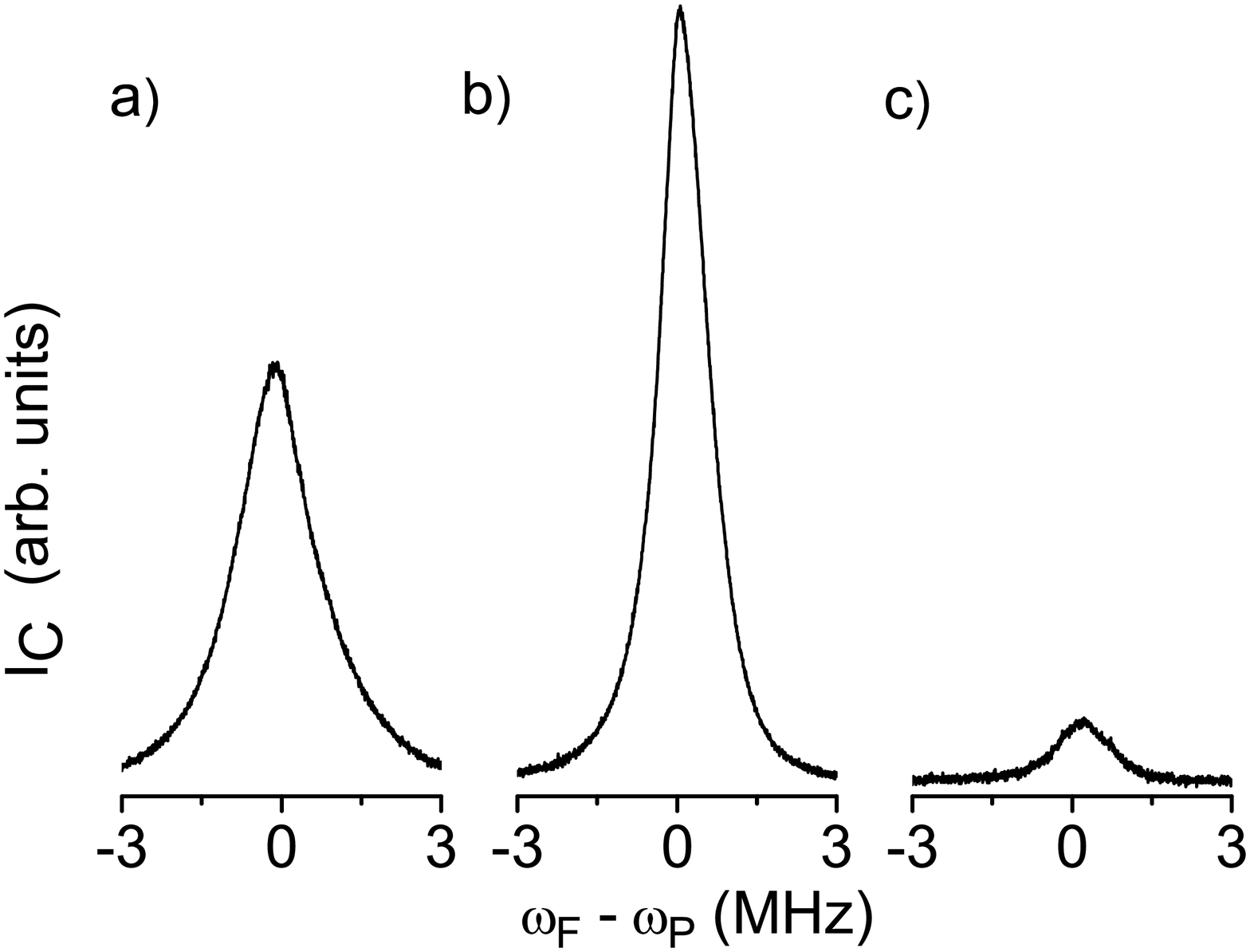}
    \caption{Intensity of conjugated beam ($I_C$) when the frequency $\omega_P$ is scanned
    around a fixed value of $\omega_F$ for a) $\omega_F=\omega_{32}$, b) $\omega_F=\omega_{32}+2\pi\times38$ MHz,
and c) $\omega_F=\omega_{CO}$. The powers used for the beamas are
$P_F=1$ mW, $P_B=0.7\,P_F$, and $P_P=150\:\mu W$. The same
vertical scale was used in all graphs.}
    \label{fig:ScanPHpower}
\end{figure}

When we increase the pump power to $P_F = 1$~mW, the relative
intensities of the conjugated signal for different fixed $\omega_F$
exhibit drastic changes (Fig.~\ref{fig:ScanPHpower}). A strong DBFWM signal occurs when $\omega_F$
is between the crossover resonance and  $\omega_{32}$, around $2\pi\times38$ MHz on the blue
side of the $F=3\rightarrow F'=2$ transition (see Fig.
\ref{fig:ScanPHpower}(b)). The signal for this frequency is larger
than the signal for the $\omega_F=\omega_{32}$ resonance (Fig.
\ref{fig:ScanPHpower}(a)), and the signal for $\omega_F=\omega_{CO}$ (Fig. \ref{fig:ScanPHpower}(c)) is strongly suppressed. The peaks at Figs.~\ref{fig:ScanPHpower}(a) and (c) also present a noticeable larger linewidth, indicating stronger power broadening at those conditions.

\begin{figure}[h]
    \centering
        \includegraphics[width=8.5cm]{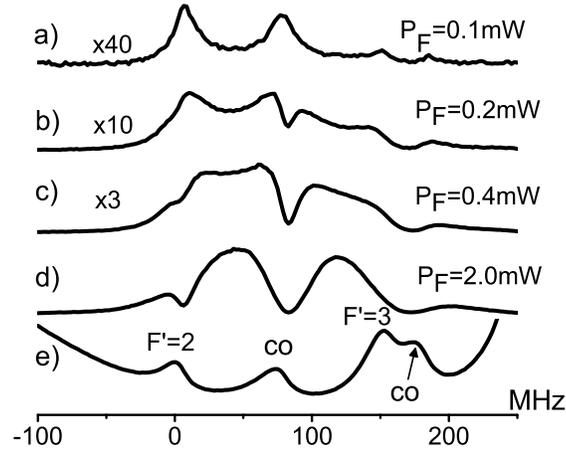}
    \caption{Intensity of the conjugated beam while scanning the
    laser frequency, keeping $\omega_F=\omega_P=\omega_B$. The pump power is equal to: a)
$100\mu W$; b) $200\mu W$; c) $400\mu W$ and d) $2000\mu W$. The probe power for curves (a)-(d) is $P_P=100\mu W$.
The saturated absorption signal used to calibrate the frequency
scan is shown in (e).}
    \label{fig:EspectroMQO}
\end{figure}

In order to investigate in detail this dependence of the DBFWM signal for different laser frequencies and power, we measured the peak height (obtained for $\omega_P = \omega_F$) as a function of $\omega_F$ throughout the Doppler profile around the $3 \rightarrow 2$ and $3 \rightarrow 3$ transitions, and for various pump powers. The results are shown in Fig.~\ref{fig:EspectroMQO}, together with the saturated absorption signal indicating the value of $\omega_F$ at each position. The condition $\omega_P = \omega_F$ was monitored by measuring a frequency beating in a fast photodiode. From this figure it is possible then to capture the broader picture of the behavior highlighted in Figs.~\ref{fig:ScanP} and~\ref{fig:ScanPHpower}. We clearly observe the two peaks of the DBFWM signal at $\omega_F=\omega_{32}$ and $\omega_F=\omega_{CO}$ for low pump power (curve (a) for $P_F=100\,\mu$W). In addition, we notice that there are also smaller peaks around the two other resonances present in the region scanned by $\omega_F$: the transition $F=3 \rightarrow F^{\prime}=3$ (at $\omega_{33}$) and the crossover for the transitions $F=3 \rightarrow F^{\prime}=2$ and $F=3 \rightarrow F^{\prime}=4$. We understand that the $F=3 \rightarrow F^{\prime}=3$ peak is largely suppressed due to optical pumping to $F=4$. The smaller height for the $F^{\prime}=2 - F^{\prime}=4$ crossover peak should result from a combination of optical pumping and the smaller strenght of the transitions to $F^{\prime}=4$.

When the pump power is increased, the power spectrum changes completely (curves (b)-(d) on Figs. \ref{fig:EspectroMQO}). The intensity of the conjugate signal ($I_C$) at the crossover resonance changes from a peak to a dip, with a similar passage from maximum to minimum occuring around the frequencies $\omega_{32}$ and $\omega_{33}$ for higher pump powers. For high pump powers we obseve now that the largest signals occur in the frequency region between the crossover and the $\omega_{32}$ and $\omega_{33}$ resonances.

\section{Theoretical model for four wave mixing process}

In order to understand the observed dependence of the DBFWM signal with both frequency and power of the pump beams, we developed a theoretical model to calculate the corresponding
DBFWM intensity lineshapes. To simplify the notation, we consider
levels $\Ket{a}$,$\Ket{b}$,$\Ket{c}$, and $\Ket{d}$ as being,
respectively, the two degenerated ground states (in our experiment
two 6$S_{1/2}$ Zeeman sublevels), and two single Zeeman excited states of levels 6P$_{3/2}$
F'=2 and 6P$_{3/2}$ F'=3. This corresponds to the simplified four-level scheme depicted in Fig.~\ref{fig:Config}, already used to explain the experimental configuration. According to the beam polarizations, the
F beam couples level $\Ket{a}$ to levels $\Ket{c}$ and $\Ket{d}$,
while the P and B beams couple level $\Ket{b}$ to $\Ket{c}$ and
$\Ket{d}$.

\subsubsection{Formalism}
We write the system Hamiltonian, in the rotating wave approximation, as:
\begin{equation}
H=H_0+V_F+V_B+V_P\,,
\end{equation}
with
\begin{eqnarray}
H_0&=&\sum_j\hbar\omega_j\KetBra{j}{j}\,,\\
V_F&=&\frac{\Omega_F\hbar}{2}e^{i\left(\omega^{at}_Ft-\vec{k}_F\cdot\vec{r}\right)}\sum_j\KetBra{a}{j}+h.c\,,\\
V_I&=&\frac{\Omega_I\hbar}{2}e^{i\left(\omega^{at}_It-\vec{k}_I\cdot\vec{r}\right)}\left(\KetBra{b}{c}-\KetBra{b}{d}\right)+h.c\,,
\end{eqnarray}
where $j=c,d$; $I=B,P$; $\hbar\omega_j$ is the energy of level
$\Ket{j}$ (levels $\Ket{a}$ and $\Ket{b}$ are considered at zero
energy); $\vec{k}_{I}$ is the wavevector of field $I$;
and h.c denotes hermitian conjugated. $\Omega_F=\frac{DE_F}{\hbar}$ and $\Omega_I=\frac{DE_I}{\hbar}$ ($I=B,P$) are the modulus of the Rabi frequencies for the various transitions. For simplicity, we have considered the electric dipole moment $D$ with the same modulus for all transitions, but with an opposite sign for the transitions from $\Ket{b}$ to $\Ket{d}$ (see Ref.~\cite{Steck} for the sign of the different dipole moments). The field frequencies
$\omega^{at}_P$ and $\omega^{at}_I$ are written in the moving
atomic frame and are related to the frequencies in
the laboratory frame by the Doppler shift: $\omega^{at}_P=\omega_P-\vec{k}\cdot\vec{v}=\omega^{at}_F$ and  $\omega^{at}_B=\omega_B+\vec{k}\cdot\vec{v}$.

We have used the density matrix formalism to calculate the
polarization terms responsible for the generation of the conjugated
beam. The evolution of the density matrix is given by the Bloch
equations:
\begin{equation}
\frac{d}{dt}\rho=-\frac{i}{\hbar}\left[H,\rho\right]+\left(\mbox{r.t.}\right)\,.
\end{equation}
To write the relaxation terms (r.t.) we chose to treat the atomic system as closed with both excited states decaying with the same rate $\Gamma$ to the ground states. We also simplify the problem by considering that the
excited state has equal probability to decay to each ground
state and that the decaying of the coherence $\rho_{ab}$ is given
by  a relaxation rate $\gamma$, which can be related to the atomic transit time in the cross sections of the beams.

Although
in our experiment all the laser beams are cw, the stationary
solutions of the Bloch equations cannot be obtained simply
imposing $\dot{\rho}_{ij}=0$, since there can be some beating on
density matrix elements involving level $\Ket{b}$ because beams B
and P can have different frequencies (in the atomic frame). To
circumvent this problem we have chosen to treat the system of
equations perturbatively in $\Omega_P$, such that the beating
between $\omega^{at}_P$ and $\omega^{at}_B$ can be neglected. We
though consider the following system of equations:
\begin{eqnarray}
\frac{d}{dt}\rho^{(0)}&=&-\frac{i}{\hbar}\left[H_0+V_F+V_B,\rho^{(0)}\right]+\left(\mbox{r.t.}\right)\,, \label{ro0}\\
\frac{d}{dt}\rho^{(P)}&=&-\frac{i}{\hbar}\left[H_0+V_F+V_B,\rho^{(P)}\right]-\frac{i}{\hbar}\left[V_P,\rho^{(0)}\right] \nonumber \\ &&+\left(\mbox{r.t.}\right)\,, \label{roP}
\end{eqnarray}
where $\rho^{(0)}$ and $\rho^{(P)}$ denotes, respectively, density matrix elements in order zero and in order one in $\Omega_P$.\\

From Eqs.~(\ref{ro0}) and~(\ref{roP}) we get two coupled systems of equations containing
rotating exponential terms such as $e^{i\left(\omega^{at}_At-\vec{k}_A\cdot\vec{r}\right)}$
(where A=F,P,B). To eliminate such terms we introduce the transformations\cite{Bloch83}:
\begin{eqnarray}
\rho^{(0)}_{ij}&=&\sum_{a,b}\sigma^{(0),(a,b)}_{ij}e^{i\left[\left(a\omega^{at}_F+b\omega^{at}_B\right)t-\left(a\vec{k}_F+b\vec{k}_B\right)\cdot\vec{r}\right]}\,, \\ \label{expansion0}
\rho^{(P)}_{ij}&=&\sum_{a,b,c}\sigma^{(P),(a,b,c)}_{ij}\times \nonumber \\ && \;\;\times e^{i\left[\left(a\omega^{at}_P+b\omega^{at}_B+c\omega^{at}_P\right)t-\left(a\vec{k}_P+b\vec{k}_B+c\vec{k}_P\right)\cdot\vec{r}\right]}\,,\label{expansion}
\end{eqnarray}
where $a,b$ are integer numbers and $c=\pm1$. Introducing such
expansion in the optical Bloch equations (Eq.~(\ref{roP})) and
looking for stationary solutions, we obtain recurrence relations
between the different $\sigma^{(P),(a,b,c)}_{ij}$ and
$\sigma^{(0),(a,b)}_{ij}$ coefficients. The conjugated beam in our
four wave mixing configuration (see Fig. \ref{fig:Config}(b)) is
generated by the optical coherence terms between the excited levels
and the ground level $\Ket{a}$ with exponential time and space
dependence such that
$I_C\propto\left|\sigma^{(P),(1,1,-1)}_{ac}+\sigma^{(P),(1,1,-1)}_{ad}\right|^2$.
We have solved numerically the derived system of recurrence
equations and calculated the DBFWM lineshapes by also numerically
integrating the corresponding optical coherence over the
Maxwellian velocity distribution of the atomic ensemble.

\subsubsection{Comparison with experimental results}

\begin{figure}[h]
    \centering
        \includegraphics[width=8.5cm]{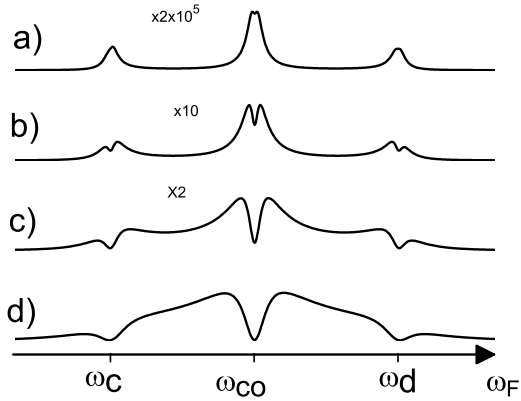}
    \caption{Calculated spectrum for the conjugated beam intensity as a function of the beam $F$
    frequency ($\omega_F$) for various Rabi frequencies of the pump beams $\Omega_F=\Omega_B$: a) $\Omega_F=0.01\Gamma$; b) $\Omega_F=0.2\Gamma$; c) $\Omega_F=0.5\Gamma$; and d) $\Omega_F=1.0\Gamma$, where $\Gamma$ is the excited levels decay rate.}
    \label{fig:Integrado}
\end{figure}

The calculated DBFWM spectrum for the generated conjugated beam is
shown in Fig. \ref{fig:Integrado} for different values of the Rabi
frequency of the pump beams.

For low pump powers the calculated conjugated
signal, presented on curve (a) of Fig.~\ref{fig:Integrado}, exhibits clear peaks for laser frequencies resonant with
transitions to the excited levels ($\omega_{c,d}$) and the
crossover resonance ($\omega_{CO}$). For higher beam intensities the peaks at $\omega_F=\omega_{c,d}$ are deformed and shifted away, while the
crossover peak splits in two peaks, with the splitting growing
with the beam intensities in curves (b) and (c), and being completely resolved in
(d), in  a such way that the intensity spectrum
shows a dip around $\omega_F=\omega_{CO}$.

By comparing to Fig.~\ref{fig:EspectroMQO}, we observe then a quite good qualitative agreement of the calculated DBFWM lineshapes with the corresponding experimental results. As for the differences between theory and experiment, the most important is the asymmetry of the experimental curves from one side to the other of the crossover resonance. This indicates that the major simplification of our model was not to consider the optical pumping to the $F=4$ ground state, which should lead to the weaking of the signal involving the $F=3 \rightarrow F^{\prime}=3$ open transition with respect to the $F=3 \rightarrow F^{\prime}=2$ closed transition. We also did not expect to have a quantitative agreement between the experimental and theoretical saturation intensities, since we do not consider the whole Zeeman structure in our model. Taking these effects into consideration, however, we notice that such good qualitative agreement indicates that we have considered in the simplified theory the essential aspects of our experiment, as for the effective four-level system and similar magnitudes of the various dipople moments. Such theory must contain, then, the key elements to explain the experimentally observed behavior.

The main point to be clarified is the reason for having dips at the different atomic resonances, and particularly at $\omega_{CO}$, once the pump power is increased. As for the crossover resonance, one possible reason could be some destructive quantum interference, such that $\sigma^{(P),(1,-1,1)}_{ac}\approx-\sigma^{(P),(1,-1,1)}_{ad}$, due to the fact that for $\omega_F=\omega_{CO}$ there are two symmetric quantum pathways for the coherence to be built (due to the two excited levels). However, by examining  the individual terms $\sigma^{(P),(1,-1,1)}_{ac}$ and
$\sigma^{(P),(1,-1,1)}_{ad}$ issued from our calculations, at $\omega_F=\omega_{CO}$ and for different atomic velocity classes, we ruled out such effect. We understand then that the explanation for the dip formation comes from some saturation effect. We envision such saturation coming in the following way. At low intensities, the three observed peaks in Fig.~\ref{fig:Integrado} comes from the matching of the two-photon resonance for the transitions induced by $F$ and $P$ with the individual, single-photon resonances of each exciting wave with different atomic levels. At high pump powers, however, the single-photon resonances at $\omega_c$ and $\omega_d$ are dislocated due to large stark shifts induced by the pump beams. Also, the different stark shifts induced by F and B destroy the two-photon resonance for the $F$ and $P$ fields. In this way, the strongest signal at high pump powers end up occuring for nonresonant frequencies, for which the stark shifts are small and at least the two-photon resonance for $F$ and $P$ is still present.

\section{Observation of intensity correlation and anti-correlation}

\begin{figure}[h]
    \centering
        \includegraphics[width=8.5cm]{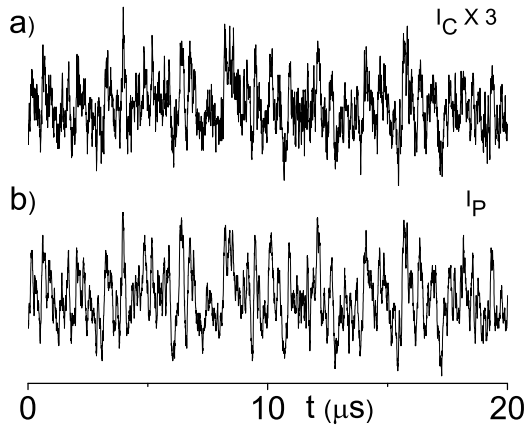}
    \caption{Time fluctuation of a) conjugate and b) probe beam intensities.}
    \label{fig:correlacaot}
\end{figure}

In order to have more insight about the physical mechanisms behind the experimental situation of the last section and to gain knowledge on the possibilities of manipulating the correlation properties of the fields, we have investigated the occurrence of classical intensity correlations between the probe and conjugated beams. We detected then, simultaneously, the transmitted probe intensity and the conjugated beam intensity with a pair of fast photodetectors connected to an also fast, 400 MHz, oscilloscope (see Fig.~\ref{fig:Setup}). The beams had similar powers at the detectors, and both the optical paths and the electronic cables carrying the photocurrents were adjusted to have the same lengths. An example of correlated behavior of the conjugated and probe photocurrents is shown in Fig.~\ref{fig:correlacaot}, which displays the signals observed on the oscilloscope as a function of time.

In order to characterize such correlations we have calculated the second
order correlation function between the two signals defined as
\cite{Scully05}:
\begin{equation}
G^{(2)}(\tau)=\frac{\left\langle\delta I_C(t)\delta I_P(t+\tau) \right\rangle}{\sqrt{\left\langle \left[\delta I_C(t)\right]^2\right\rangle\left\langle\left[ \delta I_P(t+\tau)\right]^2\right\rangle}} \label{CorrFunction}
\end{equation}
where $I_P$ denotes the intensity of the probe beam, and $\left\langle Q(t)\right\rangle=\int_t^{t+T}Q(t)dt/T$ denotes the data average over an integration time T. Such function gives quantitative information on the degree of correlation between the two fields, but it has the drawback of mixing all frequency responses and thus giving no information about the spectral behavior of the system. To acquire such information we have also calculated, thus, its corresponding Fourier transform~\cite{Cruz07}.

\begin{figure}*[h]
    \centering
        \includegraphics[width=8.5cm]{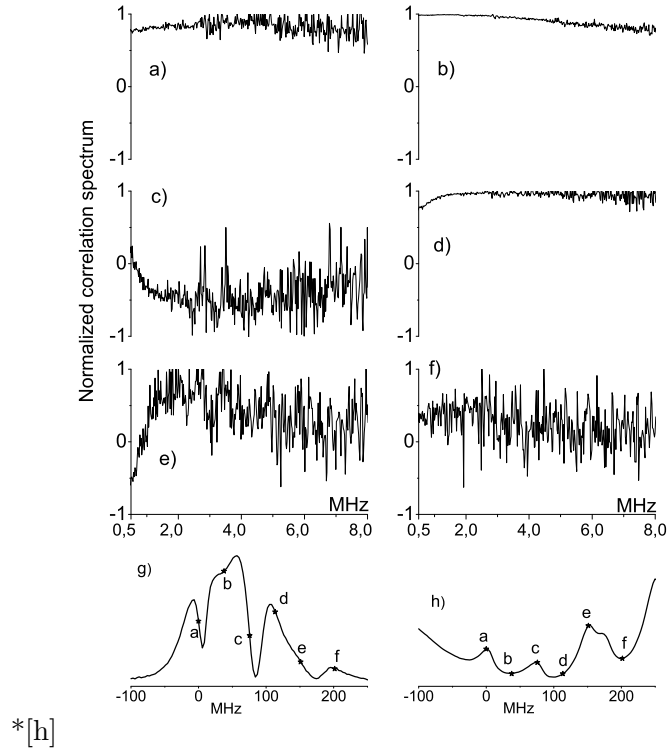}
    \caption{Normalized correlation spectrum for different laser frequencies (with $\omega_F=\omega_P=\omega_B$).
    The different frequencies analyzed are show by stars in g) and h), where conjugated intensity and
    saturated absorption signal, respectively, are shown as a function of the scanned laser frequency.}
    \label{fig:spectrum2}
\end{figure}

In Fig. \ref{fig:spectrum2} we show how the crosscorrelation
function spectrum is modified for different laser frequencies (we keep
$\omega_F=\omega_P$). The spectra shown are normalized~\cite{Cruz07}, and in Figs. \ref{fig:spectrum2}(g) and (h) we provide, respectively,  the mean intensity of the conjugated beam and the saturated absorption signal as we scan the
laser frequency, indicating the different conditions applied in (a)-(f) with respect to these quantities. We see that for most of the laser frequencies the crosscorrelation function spectrum has positive values, meaning a correlated behavior. For
instance, the correlated temporal series shown in Fig.
\ref{fig:correlacaot} correspond to the spectrum of Fig.
\ref{fig:spectrum2}(b). A different behavior, however, is observed
for Fig. \ref{fig:spectrum2}(c) for which the two beam intensity
fluctuations are anti-correlated. Such situation corresponds to
the laser frequency at the crossover resonance ($\omega_{CO}$) between $F'=2$ and $F'=3$.

\begin{figure}[h]
    \centering
        \includegraphics[width=8.5cm]{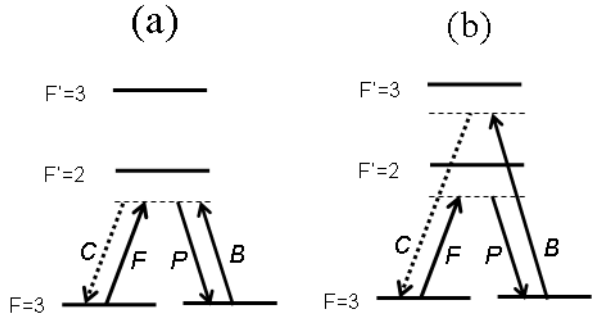}
    \caption{Field frequencies in the atomic moving frame for different laser frequencies
    and atomic velocity classes. a) $\omega_F\approx\omega_{32}$ and $v_z=0$; b) $\omega_F\approx\omega_{CO}$
    and $v_z=v_{CO}=\frac{\Delta_F}{2k}$.}
    \label{fig:corrscheme}
\end{figure}

A simple assumption related to propagation effects leading to
beams absorption by the atomic vapor can qualitatively explain the
different correlated behaviors observed for different laser
frequencies. First we should note that the parametric FWM process
leads always to intensity correlations in the fluctuation of the
probe and conjugate beams since the photons in these modes are
generated in pairs. This inherent intensity correlations should
compete with the intensity correlation or anti-correlations
induced by the differential absorption due to different atomic
velocity groups. For the laser frequency at the $F=3\rightarrow
F'=2$ transition ($\omega_F=\omega_{32}$) the main contribution to
the conjugated beam intensity comes from atoms with $v_z=0$, thus
all beams have the same frequency in the atomic frame and couple
to the same transition. When the laser frequency fluctuates (phase
fluctuation), both beams P and C (detected in our experiment)
simultaneously drive away or get closer to the atomic resonance
(see Fig. \ref{fig:corrscheme}(a)). Therefore both beams are
equally absorbed by the atomic velocity group with $v_z=0$ which
tends to increase their intensity correlation. Now if we consider
the absorption of these beams by the atomic velocity group
$v_z>0$, around $v_z=0$, for instance, for the situation depicted
in ( Fig. \ref{fig:corrscheme}(a)), the conjugate beam will be
closer to resonance than the probe beam and therefore will be more
absorbed by this group of atoms, leading to an anti-correlation in
the intensity of these beams. However, the absorption of these
beams by the symmetric velocity group $v_z<0$ will induce an
opposite intensity anti-correlation and, as the velocity
distribution function is symmetric around $v_z=0$, the two
contribution will cancel each other. Therefore, we conclude that
for the laser  frequency at the $F=3\rightarrow F'=2$ transition
($\omega_F=\omega_{32}$) the intensity fluctuation on the
transmission of the probe and conjugate beams should be always correlated
due to atomic absorption.

On the other hand, for laser frequencies at the crossover
resonance (corresponding to the spectrum of Fig.
\ref{fig:spectrum2}(c)), the main contribution to the conjugate
intensity comes from atoms with $v_z=v_{CO}=\frac{\Delta_F}{2k}$.
As before, when the laser frequency fluctuates, for instance, for
the case depicted in \ref{fig:corrscheme}(b) the absorption of the
conjugate and probe beams by this atomic velocity group will
generate correlation between these two beams. However, differently
from the previous case, the absorption of the probe and conjugate
beams by the atomic velocity groups symmetrically placed around
$v_{CO}$, which induces opposite intensity anti-correlations, will
not cancel each other since the atomic velocity distribution
around $v_{CO}$ is not symmetric, specifically
$f(v_z>\frac{\Delta_F}{2k})<f(v_z<\frac{\Delta_F}{2k})$. Thus, we
conclude that one can have an overall intensity anti-correlation
between the conjugate and the probe beam in this situation.

For laser frequencies between the crossover resonance and a
transition to an excited level, such as the situation of spectrum
\ref{fig:spectrum2}(b) where $\omega_F=\omega_{32}+2\pi\times (38$
MHz), the analysis is more complicated since the contribution to
the generated conjugate intensity comes from different velocity
groups. Such correlation measurements, thus, provide another
evidence for the different velocity classes contributing to the
signal at various conditions. In this way, it highlights once
again the possibility of choosing from a three- or a four-level
configuration for the generation of the DBFWM signal at a room
temperature vapor. It also demonstrates the possibility of
manipulating the correlation properties of the generated fields by
choosing the position of the pump beams in the Doppler profile.

\section{Conclusions}

We have investigate the four-wave mixing process in  a cesium
thermal vapor and demonstrated that this simple Doppler broadened
system can provide different atomic level configurations for
observing this nonlinear phenomenon. We also have developed a
theoretical model that reproduces reasonably well the measured FWM
spectra. In particular, the possibility of accessing a double
-$\Lambda$ level scheme with just a single laser frequency can
represent a considerable simplification for the production of
quantum correlated photons, whose generation was already
demonstrated in similar systems. Towards this direction, we have
demonstrated experimentally a strong degree of intensity
correlation between the incident probe beam and the generated
conjugated beam. Moreover, we have demonstrated that one can
switch from correlation to anti-correlation by simply changing the
frequency of the incident laser. As noted before, correlation
properties between two light beams induced in a coherent EIT media
might find application in magnetometry and atomic clocks and we
believe our results will certainly broaden this range of
applications.

ACKNOWLEDGMENT: We gratefully acknowledge P. J. Ribeiro Neto for
his contribution in the early stage of this experiment. This work
was supported by the Brazilian Agencies CNPq, CNPq/PRONEX and
FACEPE.


\begin{thebibliography}{}

\bibitem{LeBoiteux86}
S. LeBoiteux, P. Simoneau,D. Bloch, F. A. M. de Oliveira, M.
Ducloy, ``Saturation behavior of resonant degenerate 4-wave and
multiwave mixing in Doppler-Broadenend reginme - Experimental
analysis on a low-pressure Ne discharge,'' IEEE J. of Quantum
Electron. {\bf QE-22}, 1229 (1986).


\bibitem{Lukin00}
M.~D. Lukin,P.~R.~Hemmer, and M.~O.~Scully, ``Resonant nonlinear
optics in phase-coherent media,'' Adv. At. Mol. Opt. Phys. {\bf 92},
347 (2000).

\bibitem{Tabosa97}
J. W. R.Tabosa, S. S. Vianna, and F. A. M.  de Oliveira, ``Nonlinear
spectroscopy and optical phase conjugation in cold cesium atoms'',
Phys. Rev. A, {\bf 55}, 2968 (1997).


\bibitem{Lind78}
R. L. Abrams, and R. C. Lind, ``Degenerate 4-wave mixing in
absorbing media,'' Opt. Lett. {\bf 2}, 94 (1978); {\bf 3}, 205
(1978).

\bibitem{Oria89}
M. Oria, D. Bloch, M. Ficher, and M. Ducloy, ``Efficient phase
conjugation of a low-power laser diode in a short Cs vapor cell at
852 nm,'' Opt. Lett. {\bf 14}, 1082 (1989).

\bibitem{Pinard86}
M. Pinard, P. Verkerk, and G. Grynberg, ``Backward saturation in
four-wave mixing in neon: Case of cross-polarized pumps,'' Phys.
Rev. A {\bf 35}, 4679 (1986)

\bibitem{Cardoso02}
G. C. Cardoso, and J. W. R. Tabosa, ``Electromagnetically induced
gratings in a degenerate open two-level system,'' Phys. Rev. A {\bf
65}, 033803 (2002)

\bibitem{Hemmer95}
P.~R.~Hemmer, D.~P.~Katz, J.~Donioghue, M.~Cronin-Golomb,
M.~S.~Shahriar, P.~Kumar, ``Efficient low-intensity optical phase
conjugation based on coherent population trapping in sodium,'' Opt.
Lett. {\bf 20}, 982 (1995).

\bibitem{Harris06}
Pavel Kolchin, Shengwang Du, Chinmay Belthangady, G.~Y.~ Yin,and
S.~E.~Harris, ``Generation of Narrow-Bandwidth Paired Photons: Use
of a Single Driving Laser,'' Phys. Rev. Lett. {\bf 97},
113602(2006).

\bibitem{Lett07}
V.~Boyer, C.~F.~McCormick, E.~Arimondo, and P.~D.~Lett, ``Generation
of Narrow-Bandwidth Paired Photons: Use of a Single Driving Laser,''
Phys. Rev. Lett. {\bf 99}, 143601(2007).

\bibitem{Lett2007}
C.~F.~McCormick, V.~Boyer, E.~Arimondo, and P.~D.~Lett, ``Strong
relative intensity squeezing by four-wave mixing in rubidium
vapor,'' Opt. Lett. {\bf 32}, 178 (2007).

\bibitem{Harris97}
S.E. Harris, ``Electromagnetically induced transparency,'' Phys.
Today. {\bf 50}, 36 (1997).

\bibitem{Fleischhauer05}
M. Fleischhauer, A. Imamoglu, and J.~P. Marangos,
``Electromagnetically induced transparency: Optics in coherent
media,'' Rev. Mod. Phys. {\bf 77}, 633 (2005).

\bibitem{Lukin98}
M.D. Lukin, P.R. Hemmer, M. Loffler, and M. O. Scully, ``Resonant
Enhancement of Parametric Processes via Radiative Interference and
Induced Coherence,'' Phys. Rev. Lett. {\bf 81}, 2675(1998).

\bibitem{Shahriar98}
M. S. Shahriar, and P.R. Hemmer, ``Generation of squeezed states and
twin beams via non-degenerate four-wave mixing in a $\Lambda$
system,'' Opt. Communications {\bf 158}, 273(1998).

\bibitem{Garrido-Alzar03}
C.~Garrido-Alzar, L.~S.~D. Cruz, J.~G. Aguirre-G{\'o}mez, M.~F.
Santos, and P.~Nussenzveig, ``Super-Poissonian intensity
fluctuations and correlations between pump and probe fields in
Electromagnetically Induced Transparency,'' Europhys. Lett.
\textbf{61}, 485 (2003).

\bibitem{Martinelli04}
M. Martinelli, P. Valente, H. Failache, D. Felinto, L. S. Cruz, P.
Nussenzveig, and A. Lezama, ``Noise spectroscopy of nonlinear
magneto-optical resonances in Rb vapor,'' Phys. Rev. A {\bf 69},
043809(2004).

\bibitem{Scully05}
Vladimir A. Sautenkov, Yuri V. Rostovtsev, and Marlan O. Scully,
``Switching between photon-photon correlations and Raman
anticorrelations in a coherently prepared Rb vapor,'' Phys. Rev. A
{\bf 72}, 065801(2005).

\bibitem{Steck}
D.~A. Steck, Cesium D Line Data, http://steck.us/alkalidata.

\bibitem{Bloch83}
D. Bloch, and M. Ducloy, ``Theory of saturated line-shapes in
phase-conjugate emission by resonant degenerate 4-wave mixing in
Doppler-broadened 3-level systems,'' J. Opt. Soc. Am. {\bf 73}, 635 (1983); {\bf 73}, 1844 (1983).

\bibitem{Cruz07}
L.S. Cruz, D. Felinto, J.G.A. Gomez, M. Martinelli, P. Valente,
A. Lezama, P. Nussenzveig, ``Laser-noise-induced correlations and
anti-correlations in electromagnetically induced transparency,''
Euro. Phys. Journal D{\bf 41}, 531(2007).

\end{thebibliography}
\end{document}